\documentclass[preprintnumbers, floatfix, letterpaper, twocolumn,aps,prd,epsfig,nofootinbib,natbib,longbibliography]{revtex4-1}
\usepackage{bm,graphicx,dcolumn,epstopdf,epsf, latexsym,mathbbol, amssymb,amsmath,color,slashed, mathrsfs,mathcomp,simplewick}
\usepackage[center]{subfigure}
\usepackage{multirow}
\usepackage{makecell}
\usepackage[utf8]{inputenc}
\usepackage[T1]{fontenc}
\usepackage[colorlinks,linkcolor=blue,citecolor=blue,urlcolor=blue]{hyperref}
\usepackage{tensor}
\usepackage{siunitx}
\newcommand{\Lagr}{\mathcal{L}}

\newcommand*\variant[1]{\bar{#1}}

\begin{document}
\allowdisplaybreaks
 \newcommand{\bq}{\begin{equation}}
 \newcommand{\eq}{\end{equation}}
 \newcommand{\bqn}{\begin{eqnarray}}
 \newcommand{\eqn}{\end{eqnarray}}
 \newcommand{\nb}{\nonumber}
 \newcommand{\lb}{\label}
\newcommand{\f}{\frac}
\newcommand{\p}{\partial}
\newcommand{\PRL}{Phys. Rev. Lett.}
\newcommand{\PLB}{Phys. Lett. B}
\newcommand{\PRD}{Phys. Rev. D}
\newcommand{\CQG}{Class. Quantum Grav.}
\newcommand{\JCAP}{J. Cosmol. Astropart. Phys.}
\newcommand{\JHEP}{J. High. Energy. Phys.}

\preprint{YITP-21-59, IPMU21-0036}

\title{Spherically symmetric exact vacuum solutions in Einstein-aether theory}

 \author{Jacob Oost$^{1,2}$}
\email{Jacob_Oost@baylor.edu}

 \author{Shinji Mukohyama$^{3, 4}$}
\email{shinji.mukohyama@yukawa.kyoto-u.ac.jp}

 \author{Anzhong Wang$^{1,5}$\footnote{The corresponding author}}
\email{anzhong_wang@baylor.edu; corresponding author}

\affiliation{
$^{1}$ GCAP-CASPER, Physics Department, Baylor
University, Waco, TX 76798-7316, USA\\
$^{2}$ Odyssey Space Research,
1120 NASA Parkway,  Houston, TX 77058, USA \\
$^{3}$ Center for Gravitational Physics, Yukawa Institute for Theoretical Physics, Kyoto University, 606-8502, Kyoto, Japan \\
 $^{4}$ Kavli Institute for the Physics and Mathematics of the Universe (WPI), The University of Tokyo Institutes for Advanced Study,
 The University of Tokyo, Kashiwa, Chiba 277-8583, Japan\\
${}^{5}$ Institute for Theoretical Physics \& Cosmology, Zhejiang University of Technology, Hangzhou, 310023, China}

\date{\today}

\begin{abstract}
 
We study spherically symmetric spacetimes in Einstein-aether theory in three different coordinate systems, the isotropic,  
Painlev\`e-Gullstrand, and Schwarzschild coordinates,  in which the aether is always comoving,
and present both time-dependent and time-independent exact 
vacuum solutions.  In particular, in the isotropic coordinates we find a class of exact static solutions  characterized by
a single parameter $c_{14}$ in closed forms, which satisfies all the current observational constraints of the theory, and 
reduces to the Schwarzschild vacuum black hole solution in the decoupling limit ($c_{14} = 0$). However, as long
as $c_{14} \not= 0$, a marginally trapped throat with a finite non-zero radius always exists, and in one side of it  the 
spacetime is asymptotically flat, while in the other side the spacetime becomes singular within a finite proper distance 
from the throat, although the geometric area is infinitely large at the singularity.  Moreover, the singularity is a strong and 
spacetime curvature singularity, at which both of the Ricci and Kretschmann scalars become infinitely large. 
     
\end{abstract}


\maketitle

\section{Introduction}
\renewcommand{\theequation}{1.\arabic{equation}} \setcounter{equation}{0}

Recently, there has been growing interest in exploring the possibility of violations of the Lorentz invariance (LI), and~the development of new theories which feature LI-violating effects.  
Although divergences introduced by LI in quantum field theories help to motivate these explorations, LI-violations in the matter sector are tightly constrained by experiments~\cite{Kostelecky2008ts,M-L,M-Lb}.
On the other hand,  
 in the gravitational sector,  such experimental constraints  are still rather  weaker~\cite{LunarLaserRanging,SuperconductingGravimeters}, and~leave rooms for the development of theories that break 
LI,  especially in the case where the breaking is at very high energies, such as in the very early Universe.  In~particular, if~the quantization of spacetimes is a necessary feature of a full theory of  gravity, then LI must be an emergent property of low energy physics rather than a fundamental symmetry, as~it is a continuous symmetry and cannot exist in a discretized spacetime.   
Examples of theories which violate LI include Einstein-aether theory~\cite{JM01,Jacobson}   and Ho\v{r}ava gravity~\cite{Horava,SM10,Wang17}.

Einstein-aether theory (sometimes shortened as \ae-theory)  is a  vector-tensor theory that breaks LI by coupling a unit time-like vector field to the metric at every point in spacetime. It is the most general vector-tensor theory in the sense: (1) a metric theory, (2) generally covariant, (3) the aether field is unity and time-like, and~(4) the field equations are the second-order differential equations in terms of not only  the metric but also
the aether field.   It was
shown~\cite{JacobsonHorava,JacobsonHoravab} that \ae-theory  can be considered as the low-energy limit of the non-projectable Ho\v rava gravity~\cite{BPS,BPSb}.  The~theory contains three different species of gravitons,
scalar (spin-0), vector (spin-1), and~tensor (spin-2), and~each of them in principle travels at speeds not necessarily the same as the speed of light~\cite{JM04}. However, to~avoid the existence of the vacuum 
gravi-\v{C}erenkov radiation by matter, such as cosmic rays, each of them cannot be less than  the speed of light~\cite{EMS05}. Furthermore, the~gravitational wave,  GW170817, observed by the LIGO/Virgo collaboration
~\cite{GW170817}, and~the event of the gamma-ray burst GRB 170817A~\cite{GRB170817}, provides  a severe  constraint on the speed of the spin-2 mode, $- 3\times 10^{-15} < c_T -1 < 7\times 10^{-16}$.
 Nevertheless, by~properly choosing the coupling constants of the theory, it was shown that  the theory is self-consistent (such as free of ghosts and instabilities) \cite{Jacobson},
 and all the observational constraints carried out so far are satisfied~\cite{OMW18}.

In this paper, we study spherically symmetric {\em {vacuum}  
} solutions of Einstein-aether theory,   both time-dependent and time-independent, by~paying particular attention on exact solutions, solutions given analytically
in closed and explicit forms. We shall study such solutions in  three different sets of coordinate systems, namely,
{\em  the  isotropic, Painlev\`e-Gullstrand, and~ Schwarzschild coordinates}, and~present several exact solutions in closed forms.  
 In all of these studies,  we assume that the aether is at rest in the chosen coordinate~system. 

 It should be noted that  spherically symmetric {\em vacuum} spacetimes   in \ae-theory have been studied  extensively in the past couple of years both analytically~\cite{Eling2006-1,Per12,Per13,Gao2013,Dingq15,Ding16,Kai19b,Ding19,Oost2019,AA20,Chan2020,Chan2020b,Ch20,KS20,RAJH21}  and numerically~\cite{Eling2006-2, Eling2007,Tamaki2008,BS11,Enrico11,Zhu2019}.
In particular, it was shown that they can be  also formed from gravitational collapse~\cite{Garfinkle2007}. Unfortunately,  in~these studies, the~parameter space of the coupling constants of the theory 
has all been ruled out by current 
observations~\cite{OMW18}. (The only exception is the  solutions obtained by taking the limit $c_{13} \rightarrow 0$ from the ones with  $c_{14} = 0$ first found  in~\cite{Per12} in the vacuum
case, and~later generalized to the charged cases~\cite{Dingq15,Ding16,Kai19b,Ding19}, where $c_{ij} \equiv c_i + c_j$, and~$c_i\; (i = 1, 2, 3, 4)$ are the four dimensionless coupling constants of 
\ae-theory.  It is remarkable to note that such obtained solutions are the charged Schwarzschild (Reissner--Nordstrom) solutions.
Therefore, in~these limiting cases, the~aether field has no contributions to the spacetime geometry, and~can be considered either as a test field~\cite{Lin15}, or~a real time-like vector field but having
no contributions to the spacetime curvature~\cite{Zhang20}. It is equally remarkable that the aether field  remains time-like in the whole spacetime even inside 
the black holes~\cite{Lin15,Zhang20}.)  Lately, spherically symmetric BH solutions that satisfy all the observational constraints were studied numerically  in~\cite{Zhang20} and various black hole solutions were
found. It was also shown that  not only killing horizons but also a dynamical version of the universal horizons  can be formed from the gravitational collapse of realistic matter even for the coupling constants 
of the theory satisfying all the observational constraints~\cite{BMWW18}. 

Note that, due to the fact that the speeds of the spin-0 and spin-1 gravitons can, in principle, be arbitrarily large, the~boundaries of black holes in \ae-theory are no longer the locations of the
killing horizons, but~the ones of the universal horizons, which are one-way membranes for particles moving with any speeds, including the speeds that are arbitrarily large. Universal horizons 
were first proposed in~\cite{BS11} (See also~\cite{Enrico11}), and~recently have been extensively studied in~\cite{UHs,UHs1,UHs2,UHs3,UHs4,UHs5,UHs6,UHs8,UHs9,UHs10,UHs11,UHs12,UHs13,UHs14,UHs15,UHs16} (For a recent review,  see~\cite{Wang17}).

 The rest of the paper   is organized as follows: In Section~\ref{SecII} we present a brief review of \ae-theory, while in Sections~\ref{SecIII}--\ref{SecV}, we consider both static and time-dependent
 spherically  symmetric vacuum spacetimes of \ae-theory in   the isotropic,  Painlev\`e-Gullstrand, and~Schwarzschild coordinate systems, respectively, and~find various exact solutions in closed forms, and~some of which were found before but were written for the first time in closed forms. 
  The paper is ended in Section~\ref{SecVI}, in~which we summarize our main results and present some concluding remarks. There exists also an appendix, Appendix A,
  in which we present the Einstein-aether field equations in each of the three different sets of coordinate~systems.

\section{Einstein-aether theory}
\lb{SecII}
\renewcommand{\theequation}{2.\arabic{equation}} \setcounter{equation}{0}

In this paper, we consider only vacuum solutions  of the Einstein-aether theory \cite{JM01,Jacobson}, 
\begin{equation}
\lb{2.2}
    S=\frac{1}{16\pi G}\int dx^4\sqrt{-g}({\cal{R}} +\Lagr_{\mbox{\ae}}),
    \end{equation}
where ${\cal{R}}$ is the Ricci scalar and the aether Lagrangian is given by,
\begin{align}
\lb{2.3}
    \Lagr_{\mbox{\ae}}&=-\tensor{M}{^a^b_m_n}D_a u^m D_b u^n+\lambda (g_{ab}u^a u^b +1),
\end{align}
where  {$a, b = 0, 1, 2, 3$,} $D$ denotes the covariant derivative with respect to the metric $g_{ab}$, and $\lambda$ is the Lagrangian multiplier, which insures that the aether is timelike and has 
a fixed norm over the whole spacetime.  The tensor $\tensor{M}{^a^b_m_n}$ is defined as
\begin{equation}
\lb{2.4}
    \tensor{M}{^a^b_m_n}=c_1g^{ab}g_{mn}+c_2\delta ^a_m\delta ^b_n+c_3\delta ^a_n \delta ^b_m-c_4u^au^bg_{mn},
\end{equation}
where $c_i\; (i=1,2,3,4)$ are dimensionless coupling constants, as mentioned previously.   

 {Note that the above theory was first studied by Gasperini using the tetrad formalism \cite{Gasperini87}. 
On the other hand, setting $c_1 = -1/2, \; c_{13} = c_2 = c_4 = 0$, the  theory reduces to the bumblebee model,
 first proposed by Kosteleck\'y and Samuel (KS) \cite{KS89}, when the KS vector field is restricted to timelike and unity. Later, the bumblebee model was extended to the case $c_1 = (\alpha - \beta)/2, \; c_2 = \xi, \; c_3 = \xi - \alpha/2, \; c_4 = 0$, 
 where $\alpha, \beta$ and $\xi$ are three independent coupling constants, and in general the BS vector field has a vacuum expectation value, which can be timelike, null or spacelike  \cite{BK06}. }

Then, the variation of the above action with respect to $g_{ab}$ yields 
\begin{equation}
    \lb{2.1}
    G_{ab}=T^{\mbox{\ae}}_{ab}, 
\end{equation}
where $G_{ab} \equiv {\cal{R}}_{ab}-\frac{1}{2}g_{ab}{\cal{R}}$, and 
\bqn
    \lb{2.6}
    T^{\mbox{\ae}}_{ab}&\equiv& -\dfrac{1}{\sqrt{-g}}\dfrac{\lambda (\sqrt{-g}(\Lagr_{\mbox{\ae}}))}{\lambda  g^{ab}}\nb\\
    &=&  D_c\Big[\tensor{J}{^c_(_a}\tensor{u}{_b_)} + J_{(ab)}u^c-\tensor{u}{_(_b}\tensor{J}{_a_)^c}\Big]\nb\\
    && + c_1\Big[\left(D_a u_c\right)\left(D_b u^c\right) - \left(D_c u_a\right)\left(D^c u_b\right)\Big]\nb\\
    && + c_4 a_a a_b    + \lambda  u_a u_b - \frac{1}{2}  g_{ab} \tensor{J}{^d_c} \tensor{D}{_d} \tensor{u}{^c},
\eqn
where   $\tensor{J}{^a_b}$ and  $a^a$ are defined by
\bqn
    \lb{2.10}
    &&\tensor{J}{^a_b}=\tensor{M}{^a^c_b_d}\tensor{D}{_c}\tensor{u}{^d},\quad
    a^a=u^b D_b u^a.
\eqn
 
In addition, the variation of the action with respect to $u^a$ yields the aether field equations,  
\bqn 
    \lb{2.7}
   \mbox{\AE}_a&=& \frac{1}{\sqrt{-g}}\frac{\lambda  \left(\sqrt{-g} \Lagr_{\mbox{\ae}}\right)}{\lambda  u^a}\nb\\
    &=& \ D_{\alpha} J^{\alpha}_{\;\;\;\mu} + c_4 a_{\alpha} D_{\mu}u^{\alpha} + \lambda u_{\mu} = 0,
\eqn 
while its variation   with respect to $\lambda$ gives, 
\bq
\lb{2.7b}
u^a u_a = -1.
\eq
From Eqs.(\ref{2.7}) and (\ref{2.7b}) we find that 
\begin{equation}
    \lb{2.12}
    \lambda = u_b D_a \tensor{J}{^a^b} + c_4 a^2.
\end{equation}

As mentioned above, the theory in general allows three different species of gravitons, spin-0, spin-1, and spin-2, and each of them move in principle with different speeds, given, respectively, by \cite{JM04},
\bqn 
\lb{2.12a}
c_S^2 & = & \frac{c_{123}(2-c_{14})}{c_{14}(1-c_{13}) (2+c_{13} + 3c_2)}\,,\nonumber\\
 c_V^2 & = & \frac{2c_1 -c_{13} (2c_1-c_{13})}{2c_{14}(1-c_{13})}\,,\nonumber\\
 c_T^2 & = & \frac{1}{1-c_{13}},
\eqn
where   {$c_{ij} \equiv c_i + c_j,  \; c_{ijk} \equiv c_i + c_j + c_k$,} and $c_{S,V, T}$ represent the speeds of the spin-0, spin-1, and spin-2 gravitons, respectively. 

The most recent observational constraints on the coupling constants $c_i$, in light of the LIGO/Virgo gravitational wave detection GW170817 \cite{GW170817} and its concurrent gamma-ray burst GRB170817A \cite{GRB170817}, were found in \cite{OMW18}, together with the self-consistent conditions, such as the absence of ghosts and instability \cite{Jacobson}. Depending on the values of $c_{14}$, the constraints 
can be divided into three different bands, and are given, respectively, by  \cite{OMW18}, 
\bqn
\lb{2.15a}
&& (i) \;\; 0 \leq   c_{14}  \leq 2\times 10^{-7}: \nb\\
&&  ~~~~~~~  c_{14} \leq c_2 \leq 0.095,\\
\lb{2.15b}
&& (ii) \;\; 2\times 10^{-7} \leq  c_{14}  \leq 2\times 10^{-6}:\nb\\
&& ~~~~~~~ c_{14} \leq c_2\leq 0.095,\nb\\
&& ~~~~~~~ - 10^{-7} \le  \frac{c_{14}\left(c_{14} + 2c_2c_{14} - c_2\right)}{c_2\left(2-c_{14}\right)} \le 10^{-7},~~~~\\
\lb{2.15c}
&& (iii) \;\; 2\times 10^{-6}\leq  c_{14}  \leq 2.5\times 10^{-5}: \nb\\
&& ~~~~~~~ 0 \leq c_2 - c_{14} \leq c_2 \times 10^{-7}.  
\eqn
Therefore, each of the three parameters, $c_2, \; c_{14}$ and $c_{13}$ are restricted, respectively, to the ranges, 
\begin{align}
    \lb{2.15}
    &0<c_2\lesssim 0.095,\\
    \lb{2.16}
    &0<c_{14}\lesssim \num{2.5e-5},\\
    \lb{2.17}
    &\mid c_{13}\mid \; \lesssim \num{10e-15}.
\end{align}

\section{Spherically symmetric  Spacetimes in Isotropic Coordinates}
\lb{SecIII}
\renewcommand{\theequation}{3.\arabic{equation}} \setcounter{equation}{0}

 \subsection{Spherically Symmetric Spacetimes}

The general form for a spherically-symmetric metric can be written as,
\bqn
    \lb{1.3}
    ds^2&=&g_{AB} dx^Adx^B+R^2 d\Omega^2\nb\\
    &=& - {\cal{N}}^2 dt^2 + {\cal{B}}^2\left(dr + {\cal{N}}^r dt\right)^2 +R^2d\Omega^2,
\eqn
where  ${\cal{N}}$,  ${\cal{B}}$,  ${\cal{N}}^r$, and $R$ are functions of $t$ and $r$ only, $x^{\mu} = (t, r, \theta, \phi)$, and  $d\Omega^2 \equiv d\theta^2 + \sin^2\theta d\phi^2$.   This metric clearly is invariant under the coordinate transformations, 
\bqn 
    \lb{1.4}
    &&t=f(\variant{t},\variant{r}), \quad 
    r=g(\variant{t},\variant{r}), 
    \eqn
where $f$ and $g$  are arbitrary functions of their indicated arguments. By properly choosing these functions, we are able to fix two of the four arbitrary functions  ${\cal{N}}$,  ${\cal{B}}$,  ${\cal{N}}^r$, and $R$.

In this section, we shall use the gauge freedom (\ref{1.4}) to set,  
\bqn 
    \lb{4.0a}
     g_{rr} = R(t, r),  \quad g_{tr} = 0, 
 \eqn 
so that the metric (\ref{1.3}) takes the form,   
\begin{equation}
    \lb{4.1}
    ds^2=-e^{2\mu(r,t)}dt^2+e^{2\nu(t,r)}d\sigma^2,
\end{equation}
Where $d\sigma^2$ is the spatial part of the metric, defined as,
\begin{equation}
    \lb{4.1a}
    d\sigma^2 \equiv dr^2+r^2\left(d\theta^2+\sin{\theta}^2d\phi^2\right).
\end{equation}
Then,  the comoving  aether \footnote{ {Here ``comoving aether" means that the aether field is at rest in the chosen coordinates, so it has only the timelike
component, while its spatial components vanish identically, i.e.,  $u^{i} = 0\; (i = 1, 2, 3)$. When the spacetime is static, it aligns with the timelike Killing vector, $\xi^{\mu} = \delta^{\mu}_t$.}} is  given by 
\begin{equation}
    \lb{4.2}
    u^{a}=e^{-\mu} \delta _t^{a}.
\end{equation}

To write down the field equations, we find convenient first to introduce the constant  $\alpha$ and the function $\Sigma$ as, 
\bqn
    \lb{4.2a}
    \alpha^2 &\equiv& 3\left(1+\frac{3 c_2+c_{13}}{2} \right),\\
    \lb{4.2b}
    \Sigma &\equiv& 3\dot{\nu}^2+2\ddot{\nu}-2\dot{\mu}\dot{\nu}.
\eqn

Then, the non-vanishing equation for the aether dynamics is,
\begin{equation}
\lb{4.3}
    0=(3 c_2+c_{13}+c_{14})\mu'\dot{\nu}+c_{14}\dot{\mu}'-\beta\dot{\nu}',
\end{equation}
Where $\beta \equiv 3c_2+c_{13}$. The non-vanishing components of  $G_{\mu\nu}$ and $T^{\mbox{\ae}}_{\mu\nu}$ are given by
Eqs.(\ref{B.2}) - (\ref{B.11}), from which we find that currently there are four non-trivial equations, given, respectively, by 
the $tt$, $tr$, $rr$, $\theta\theta$ components,   
\bqn
\lb{4.4}
    && \alpha^2\dot{\nu}^2 e^{2\nu} =e^{2\mu}\Bigg[c_{14}\left(\frac{\mu'^2}{2}+\mu'\nu'+\mu''+2\frac{\mu'}{r} \right)\nb\\
    &&~~~~~~~~~~~~~~~~~~~~~ +\nu'^2+2\nu''+4\frac{\nu'}{r}\Bigg],\\
    \lb{4.5}
    &&c_{14}(\dot{\mu'}+\mu'\dot{\nu})=2(\mu'\dot{\nu}-\dot{\nu'}), \\
    \lb{4.6}
    &&\frac{\alpha^2}{3} e^{2\nu} \Sigma =e^{2\mu}\Bigg[\nu'^2+2\mu'\nu'+\frac{2}{r}(\mu'+\nu')\nb\\
    &&~~~~~~~~~~~~~~~~~~~~~ +c_{14}\frac{\mu'^2}{2}\Bigg], \\
    \lb{4.7}
    &&\frac{\alpha^2}{3}e^{2\nu} \Sigma =e^{2\mu}\Bigg[\mu'^2+\mu''+\nu''+\frac{\mu'+\nu'}{r}\nb\\
    &&~~~~~~~~~~~~~~~~~~~~~ -c_{14}\frac{\mu'^2}{2}\Bigg]. 
\eqn

\subsection{Time-independent Solutions}

With no time-dependence, the five equations are reduced to three.  Then, from the $tt$ and $\theta\theta$ equations,
we find 
\begin{equation}
    \lb{4.static1}
    0=f''+f'^2+\frac{3}{r}f',
\end{equation}
where
\begin{equation}
    \lb{4.static2}
    f\equiv \mu+\nu.
\end{equation}
To solve Eq.(\ref{4.static1}) we first divide both sides of the equation by $f'$ and then integrate it, leading to, 
\begin{equation}
    \lb{4.static3}
    \ln{(L_0f')}=-f-3\ln{\left(\frac{r}{r_0}\right)},
\end{equation}
where $L_0$ and $r_0$ are the integration constants with dimensions of length.  
Eq.(\ref{4.static3}) has the general solutions,  
\begin{equation}
    \lb{4.static5}
    f = \ln{\left(f_0\left(1-\frac{r_0^2}{r^2} \right) \right)},
\end{equation}
where $f_0\equiv r_0/2L_0$ is a dimensionless constant.  Next we subtract the $rr$ equation from the $tt$ one,   leading to, 
\begin{equation}
    \lb{4.static6}
    2\nu''+2\frac{\nu'}{r}=(2-c_{14})\mu'\nu'+(2-2c_{14})\frac{\mu'}{r}-c_{14}\mu''.
\end{equation}
Now, from the $\theta\theta$ equation, we find, 
\begin{equation}
    \lb{4.static7}
    2\nu''+2\frac{\nu'}{r}=-2\mu'^2-2\mu''-2\frac{\mu'}{r}+c_{14}\mu'^2.
\end{equation}
The combination of  Eqs.(\ref{4.static6}) and (\ref{4.static7}) yields, 
\begin{equation}
    \lb{4.static8}
    0=(c_{14}-2)\left[\mu'^2+\mu''-2\frac{\mu'}{r}+\mu'\nu' \right].
\end{equation}
Note that the observational constraints lead to Eq.(\ref{2.16}), from which we can see that $c_{14} - 2\neq 0$ always holds. Therefore,
the above equation yields,
\begin{equation}
    \lb{4.static16}
    0=\mu'^2+\mu''+2\frac{\mu'}{r}+\mu'\nu',
\end{equation}
which has the solution,  
\begin{equation}
    \lb{4.static17}
    f =-\ln{\left(\frac{f_0r_0q}{\mu'r^2} \right)},
\end{equation}
where  $q$ is an arbitrary dimensionless constant. Then, combining   Eqs.(\ref{4.static5}) and (\ref{4.static17}), we find,  
\begin{equation}
    \lb{4.static18}
    \mu'=\frac{r_0 q}{r^2-r_0^2},
\end{equation}
which  yields,  
\begin{equation}
    \lb{4.static19}
    \mu=\frac{q}{2}\ln{\left(U_0\frac{r-r_0}{r+r_0} \right)},
\end{equation}
where $U_0$ is a dimensionless constant.  We can now solve for $\nu$ by using Eqs.(\ref{4.static5}) and  (\ref{4.static19}), and find,   
\begin{equation}
    \lb{4.static20}
    \nu=\ln{\left[\frac{f_0}{U_0}\left(1-\frac{r_0^2}{r^2} \right) \left(\frac{r+r_0}{r-r_0} \right)^{\dfrac{q}{2}} \right]}.
\end{equation}
These solutions for $\mu$ and $\nu$ solve the field equations exactly provided that q is given by,
\begin{equation}
    \lb{4.static21}
    q \equiv 2\sqrt{\dfrac{2}{2-c_{14}}}.   
\end{equation}
Since $0 \le c_{14} \le 2.5 \times 10^{-5}$, we find that 
\begin{equation}
    \lb{4.static21a}
    2 \le q \lesssim 2\left(1 + 1.25 \times 10^{-5}\right).   
\end{equation} 
So, the spacetime is given by,
\bqn
    \lb{4.static22}
    ds^2&=&\frac{f_0^2}{U_0^2} \Bigg\{-\frac{U_0^{2+q}}{f_0^2} \left(\frac{r-r_0}{r+r_0} \right)^qdt^2\nb\\
    && +\left(1-\frac{r_0^2}{r^2} \right)^2\left(\frac{r+r_0}{r-r_0} \right)^qd\sigma^2\Bigg\}.
\eqn
Rescaling $t$ we can set the factor $U_0^{2+q}/f_0^2 = 1$, so the above metric  {takes} the form $ds^2 = \left({f_0^2}/{U_0^2}\right) d\bar s^2$.  
Since $ds^2$ and $d\bar s^2$ are conformally related by a constant, the spacetimes described by them have the same properties. Therefore, without loss of the generality,
we can always  set $f_0 = U_0 = 1$. 

On the other hand, to see the meaning of $r_0$, let us consider the Schwarzschild metric in  the isotropic coordinates, which  is given by, 
\begin{equation}
    \lb{4.static23}
    ds^2=-\left(\frac{1-\frac{m}{2r}}{1+\frac{m}{2r}} \right)^2 dt^2+\left(1+\frac{m}{2r} \right)^4 d^2\sigma,
\end{equation}
where $d^2\sigma \equiv  dr^2 + r^2d^2\Omega$, as noticed previously. In the $c_{14}\rightarrow 0$ limit, $q\rightarrow 2$, so the spacetime given by Eq.(\ref{4.static22}) 
does indeed reduce to the isotropic Schwarzschild solution given by Eq.(\ref{4.static23}),
 provided that
\bq
    f_0=1,\quad
    U_0=1,\quad
    r_0=\frac{m}{2},
\eq
for which the metric (\ref{4.static22}) takes the form, 
\bqn
    \lb{4.static24}
    ds^2=-\left(\frac{1-\frac{m}{2r}}{1+\frac{m}{2r}} \right)^qdt^2
     +\left(1-\frac{m^2}{4r^2} \right)^2\left(\frac{1+\frac{m}{2r}}{1-\frac{m}{2r}} \right)^q d^2\sigma,\nb\\
\eqn
where $q$ is given by Eq.(\ref{4.static21}) and $r\geq m/2$ (which is also true for the Schwarzschild solution in isotropic coordinates).

\begin{figure}[htb]
 	\includegraphics[width=\columnwidth]{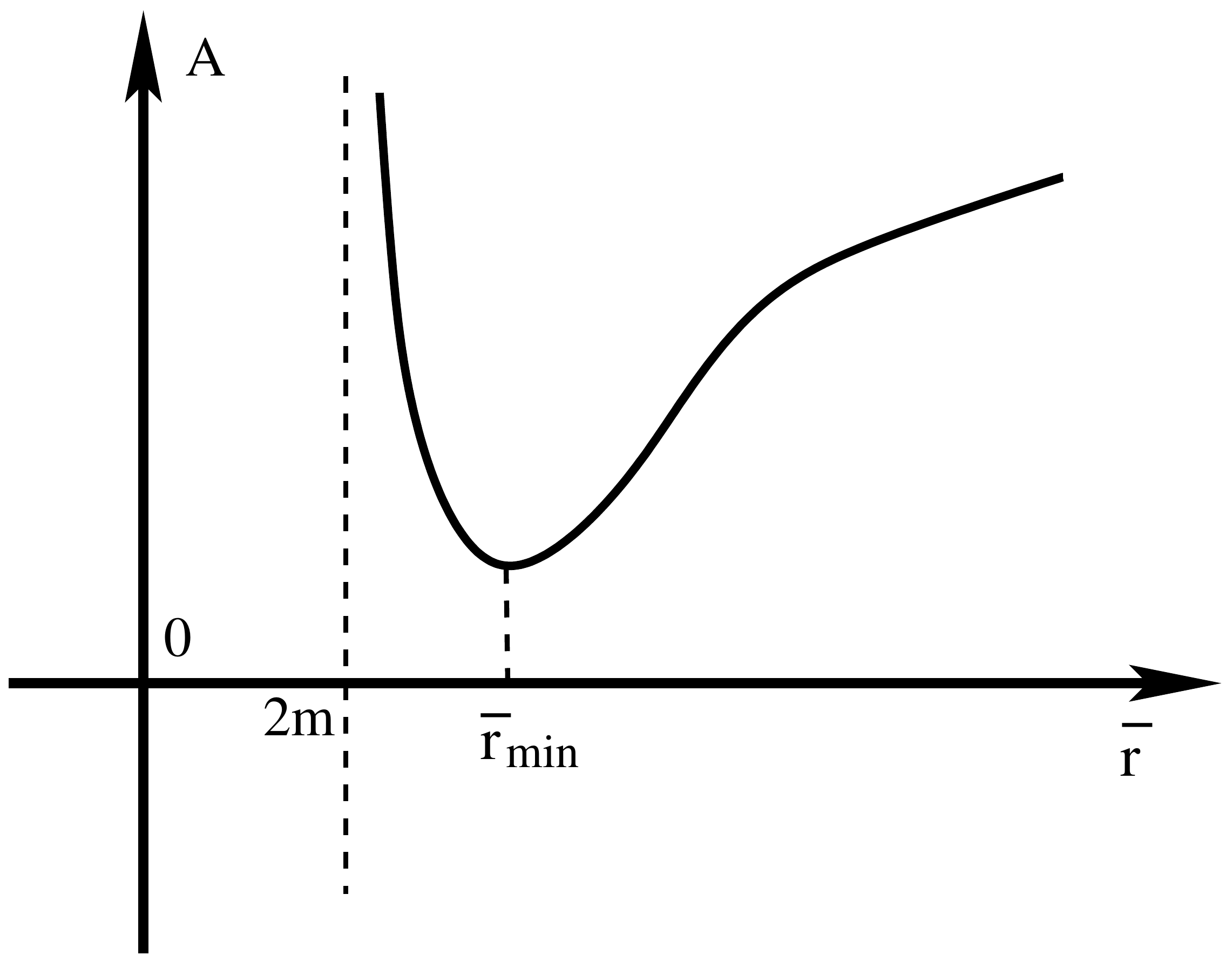} 
 	\caption{The geometric area $A$ of the 2-dimensional spheres $t, \bar{r} = $ Constant vs the radial coordinate $\bar{r}$  for the spacetime described by the metric (\ref{Static_4.static26}) with $c_{14} \not = 0$, where $A$ is given  by Eq.(\ref{Static_4.static35}).  The point 
	$\bar{r} = \bar{r} _{\text{min}}$ defined by  Eqs.~(\ref{Static_4.static25}) and (\ref{Static_4.static36}) is the minimal surface for the area $A$ - the throat. The area $A$ becomes infinitely large at both $\bar{r} = 2m$ and $\bar{r} = \infty$, 
	and the spacetime is singular at $\bar{r} = 2m$, but asymptotically flat as $\bar{r} \rightarrow \infty$.} 
 	\label{fig1}
 \end{figure}

The spacetime given by Eq.(\ref{4.static24}) has curvature singularities at $r=\frac{m}{2}$ and at $r=0$.
  Both are curvature singularities as can be seen by considering the Ricci scalar. However,  this is easier to see in  {a} coordinate system similar to the Schwarzschild form. Consider the coordinate transformation:
\begin{equation}
    \lb{Static_4.static25}
    \bar{r}=r\left(1+\frac{m}{2r}\right)^2,
\end{equation}
upon which the metric becomes
\begin{align}
    \lb{Static_4.static26}
    ds^2&=-\left( 1-\frac{2m}{\bar{r}}\right)^{q/2}dt^2+\left( 1-\frac{2m}{\bar{r}}\right)^{-q/2}d\bar{r}^2\nb\\
    &~~~~+\left( 1-\frac{2m}{\bar{r}}\right)^{1-q/2}\bar{r}^2d\Omega^2.
\end{align}
Then,  the Ricci scalar is given by,
\begin{equation}
    \lb{Static_4.static28}
    {\cal{R}}    = \frac{m^2(4-q^2)}{2\bar{r}^4}\left( 1-\frac{2m}{\bar{r}}\right)^{\frac{q}{2}-2},
\end{equation}
and the Kretschmann scalar is given by
\bqn
    \lb{Static_4.static29}
    {\cal{K}} &\equiv& R_{abcd}R^{abcd} \nb\\
    &=&  \frac{m^2}{4\bar{r}^4}\left( 1-\frac{2m}{\bar{r}}\right)^{q-4}(a\bar{r}^2+b\bar{r}+d),
\eqn
where
\bqn 
    \lb{Static_4.static30}
    &&a=48q^2, \;\;\; b=-32mq(q^2+3q+2),\nb\\
    &&d=m^2(2+q)^2(7q^2+4q+12).
\eqn 
Obviously both the Ricci and Kretschmann scalars have curvature singularities at the origin, and upon carefully taking the limit when $\bar{r}$ approaches $2m$ we see that there are curvature singularities at $\bar{r}=2m$ as well.  When $c_{14}$ is set to zero they reduce to the correct values for the Schwarzschild solution's Ricci and Kretschmann scalars (expressed in the Schwarzschild coordinates).  As can be seen from Eq.(\ref{Static_4.static26}) the area of a sphere centered on the origin is given by,
\begin{equation}
    \lb{Static_4.static35}
     {A}=4\pi \bar{r}^2\left(\frac{\bar{r}}{\bar{r}-2m}\right)^{q/2 - 1}.
\end{equation}
When $c_{14}=0$ we have $q = 2$, and then $\bar{r}$ becomes the areal radial coordinate and a sphere with coordinate  radius $\bar{r}=2m$ has the area $4\pi \bar{r}^2$ as expected.  However,
when $c_{14} \not= 0$, we have $q \ge 2$, and  the area of a sphere 
is well defined only for $\bar{r} \ge 2m$  [cf. Eq.(\ref{4.static21a})], and becomes  infinitely large at both $\bar{r}=2m$ and $\bar{r}=\infty$.  This shows that while the spacetime of Eqs.(\ref{4.static24}) and (\ref{Static_4.static26}) do approach the Schwarzschild solution as $c_{14}$ approaches zero, the approach is not completely continuous.  In particular, as long as $c_{14} \not= 0$, the
 areal radius always reaches a minimum at,
\begin{eqnarray}
    \lb{Static_4.static36}
    \bar{r} _{\text{min}}=2m\left(\frac{2+q}{4}\right) \ge 2m,
\end{eqnarray}
which serves as a throat and smoothly connects the two regions, $\bar{r} \in (2m,  \bar{r} _{\text{min}}]$ and $\bar{r} \in [\bar{r} _{\text{min}}, \infty)$, as shown 
 schematically in  Fig. \ref{fig1}, where the point 
	$\bar{r} = \bar{r} _{\text{min}}$ defined by  Eq.(\ref{Static_4.static25}) denotes the location of the throat. The spacetime is asymptotically flat as $\bar{r} \rightarrow \infty$, and 
  the proper radial distance from the throat to $\bar{r} = \infty$ is infinitely large (so is the geometric area $A$).  
  However, despite the fact that $A$ also becomes infinitely large at $\bar{r} = 2m$, the proper radial distance between the throat and $\bar{r} = 2m$ 
  is finite, 
  \bq
  \lb{Static_4.static36b}
  L \equiv \left|\int_{\bar{r}_{\text{min}}}^{\bar{r}}{\left(\frac{\bar{r}}{\bar{r} - 2m}\right)^{q/4}d\bar{r}}\right|
  = \begin{cases}
  \infty, & \bar{r} \rightarrow \infty, \cr
  {\text{finite}}, & \bar{r} \rightarrow 2m. \cr
  \end{cases}
  \eq
 {It should be noted that the proper radial distance between the throat and the singularity $\bar{r} = 2m$ becomes infinite for $3/2 < c_{14} < 2$  \cite{Eling2006-1}, for which we have $q > 4$.  However,
this violates the observational constraints given by Eq.(\ref{4.static21a}).}

To see  the nature of the curvature singularity located at $\bar{r} = 2m$, let us consider the expansions of null geodesics \cite{NumericalRelativity}.  Let the metric of Eq.(\ref{Static_4.static26}) be written as,
\begin{equation}
    \lb{Static_4.static36*}
    ds^2=-u_au_b+s_as_b+m_{ab}, 
\end{equation}
Where $u_a$ is given by Eq.(\ref{4.2}), $s_a=e^\nu\delta^r_a$ and $m_{ab}$ is the 2-dimensional metric induced on a closed, 2-dimensional, spacelike hypersurface (which is conformal to the 2-sphere metric).  Then, let the outgoing/ingoing null geodesics have the tangent vectors  {$k^{\pm}_a$,    given by,
\bq
    \lb{Static_4.static37}
    k^{\pm}_a =\frac{1}{\sqrt{2}}(u_a\pm s_a),
\eq
we find that the expansions of outgoing/ingoing null geodesics are given by
\begin{equation}
    \lb{Static_4.static38}
    \Theta_{\pm} \equiv m^{ab}\nabla_a k^{\pm}_b=\pm\frac{\sqrt{2}\left(1-\frac{2m}{\bar{r}}\right)^{q/4}}{\bar{r}\left(\bar{r} - 2m\right)}\left(\bar{r} - \bar{r} _{\text{min}}\right),
\end{equation}
which all vanish  at $\bar{r} _{\text{min}}$. However, across the throat the outgoing/ingoing null geodesics exchange their roles, so that
  $\Theta_{+} \Theta_{-} < 0$ holds in both sides of the throat. As an immediate  result,   no trapped regions exist, and the throat is only marginally trapped \cite{HE73}.}

It should be noted that  Eling and Jacobson studied the static aether case in the Schwarzschild coordinates, and found the general solutions (but given implicitly) \cite{Eling2006-1}, as to be shown explicitly in Sec. \ref{SecV}. So, the above solutions must be the same ones. In addition,    they  also found that the minimal 2-sphere, $\bar r = \bar{r} _{\text{min}}$, does not sit at a Killing horizon,  { and that
the singularity at $\bar{r} = 2m$ is actually a null singularity [cf.  \cite{Eling2006-1}, Fig.2]. In addition, these solutions with $c_{123} \not= 0$ are stable against small spherically symmetric perturbations \cite{Seifert07,Jacobson}.}

\subsection{Time-dependent Solutions}

If we consider solutions such that $e^\mu$ and $e^\nu$ are separable in $t$ and $r$, then we seek solutions of the form, 
\begin{align}
    \lb{4.8}
    \mu(r,t)&=\mu_0(r)+\mu_1(t), \\
    \lb{4.9}
    \nu(r,t)&=\nu_0(r)+\nu_1(t),
\end{align}
so that all mixed-partial derivatives of $\mu$ and $\nu$ are zero. However,   redefining   the time coordinate $t$  by $t'$,
\begin{equation}
    \lb{4.10}
    t'\equiv \int e^{2\mu_1(t)}dt,
\end{equation}
we can see that, without loss of generality, we can set $\mu_1 = 0$, and look for solutions of the form,  
\begin{align}
    \lb{4.11}
    \mu(r,t)&=\mu_0(r), \\
    \lb{4.12}
    \nu(r,t)&=\nu_0(r)+\nu_1(t).
\end{align}
  If $\dot{\nu}=0$ then the equations of motion reduce to the static case, so we assume that $\dot{\nu}\neq 0$.  In this case,  the $tr$ and aether equations reduce to, 
\bqn
    \lb{4.13}
    &&c_{14}\mu'\dot{\nu}=2\mu'\dot{\nu},\\
    \lb{4.14}
    &&c_{14}\mu'\dot{\nu}=-\beta\mu'\dot{\nu}.
\eqn
Thus, there are three possibilities, 
\bq
    (i)~~c_{14}=2=-\beta; \quad 
    (ii)~~\dot{\nu}=0;
    (iii)~~\mu'=0, \quad
\eq
where Case $(i)$ is excluded by observations and Case $(ii)$ is the static case, just studied in the last subsection. So, in the following we shall need only to consider the last case, $\mu' = 0$. 

 When $\mu'=0$,   the three relevant equations are the ones of the $tt$, $rr$, and $\theta\theta$ components, given by, 
\begin{eqnarray}
    \lb{4.15}
    &&\alpha^2 \dot{\nu_1}^2 e^{2\nu_1}=e^{2\mu_0-2\nu_0}\left(\nu_0'^2+2\nu_0''+4\frac{\nu_0'}{r} \right),\\
    \lb{4.16}
    && \alpha^2 e^{2\nu_1}\left(\dot{\nu_1}^2+\frac{2}{3}\ddot{\nu_1} \right) =e^{2\mu_0-2\nu_0}\left(\nu_0'^2+2\frac{\nu_0'}{r} \right), ~~~~\\
    \lb{4.17}
    &&\alpha^2 e^{2\nu_1}\left(\dot{\nu_1}^2+\frac{2}{3}\ddot{\nu_1} \right) =e^{2\mu_0-2\nu_0}\left(\nu_0''+\frac{\nu_0'}{r} \right).
\end{eqnarray}
Note that for each equation, the left-hand side is $t$-dependent and the right-hand side is $r$-dependent, thus both sides must be equal to the same constant.  Setting  
\begin{eqnarray}
    \lb{4.18}
    &&K_0^2\equiv \alpha^2 e^{2\nu_1}\dot{\nu_1}^2,\\
    \lb{4.19}
    &&K_1^2\equiv \alpha^2 e^{2\nu_1}\left(\dot{\nu_1}^2+\frac{2}{3}\ddot{\nu_1} \right),
\end{eqnarray}
from Eqs.(\ref{4.16}) and (\ref{4.17}) we have
\begin{equation}
    \lb{4.20}
    \nu_0''=\nu_0'^2+\frac{\nu_0'}{r},
\end{equation}
and thus it can be shown that $K_0^2=K_1^2/3$.  Eq.(\ref{4.20}) has the general solution, 
\begin{equation}
    \lb{4.25}
    \nu_0(r)=\ln{\left(\frac{r_1}{r^2-r_0^2} \right)}, 
\end{equation}
where $r_1$ and $r_0$ are integration constants. Next we solve for $\nu_1(t)$ using Eq.(\ref{4.18}),
\begin{equation}
    \lb{4.26}
    e^{2\nu_1}\alpha^2\dot{\nu_1}^2=K_0^2,
\end{equation}
which has the solution,  
\begin{equation}
    \lb{4.27}
    \nu_1(t)=\ln{\left[\frac{K_0}{\alpha}\left(t-t_0\right) \right]},
\end{equation}
where $t_0$ is an integration constant, and   $K_0 \not= 0$.  It is straightforward to show that the solutions given by Eqs.(\ref{4.25})-(\ref{4.27}) solve the field equations (\ref{4.15})-(\ref{4.17}) provided that, 
\begin{equation}
    \lb{4.28}
    \frac{K_0^2 r_1^2}{\alpha}=12 r_0^2 e^{2\mu_0}.
\end{equation}
Then, the final solutions for $\mu(t,r)$ and $\nu(t,r)$ can be expressed as, 
\begin{eqnarray}
    \lb{4.29}
    &&\mu(t,r)=\mu_0,\\
    \lb{4.30}
    &&\nu(t,r)=\ln{\left[\sqrt{\frac{12}{\alpha}}\frac{r_0(t-t_0)}{r^2-r_0^2} \right]}+\mu_0.
\end{eqnarray}
Then, using the gauge freedom $\bar t = a t + b$, where $a$ and $b$ are constants, we can always set $t_0 = \mu_0 = 0$,  so that the metric takes the form, 
\begin{equation}
    \lb{4.31}
    ds^2=-dt^2+ \frac{12r_0^2t^2}{\alpha (r^2-r_0^2)^2}\left(dr^2+r^2d^2\Omega \right).
\end{equation}

It can be shown that the spacetime described by the above metric is conformally flat, that is, the Weyl tensor vanishes identically, and the spacetime is singular at $t = 0$, as can be seen from the Ricci and Kretschmann  scalars, which are
now given by, 
\begin{equation}
    \lb{4.32}
    R=-\frac{3\beta}{t^2},\;\;\;  K=\frac{1}{t^4}\left(\frac{4}{3}\alpha^2-8\alpha+12 \right), 
\end{equation}
where $\beta=3c_2+c_{13}$, as defined previously.

To study this solution further, let us consider the energy conditions. We define a timelike vector field   $t^{\alpha}$ in the ($t, r$)-plane,  
\begin{eqnarray}
    \lb{4.34}
    t^{\alpha}=A\delta^{\alpha}_t+B\delta^{\alpha}_r,\quad A^2=v^2+B^2e^{2\nu}, 
 \end{eqnarray}
from which we find that $t^{\alpha} t_{\alpha}=-v^2$,  where $v$ is an arbitrary non-vanishing real function of $x^{\alpha}$.  
A stress-energy tensor that obeys the weak energy condition ensures that all observers following timelike trajectories will see only positive energy density \cite{HE73}, that is,  
\begin{equation}
    \lb{4.37}
    T^{\mbox{\ae}}_{\alpha\beta}t^{\alpha}t^{\beta}\geq 0.
\end{equation}
However, for the spacetime of Eq.(\ref{4.31}) we have, 
\begin{equation}
    \lb{4.38}
    T^{\mbox{\ae}}_{\alpha\beta}t^{\alpha}t^{\beta} =-3\beta\left[\frac{6B^2r_0^2}{\alpha(r^2-r_0^2)^2}+\frac{v^2}{2t^2}\right],
\end{equation} 
which is always non-positive. 
Thus,  the aether field in the current case always   violates the weak energy condition.

A stress-energy tensor that obeys the strong energy condition ensures that gravity will always be attractive, which is equivalent to require \cite{HE73},  
\begin{equation}
    \lb{4.40}
    T^{\mbox{\ae}}_{\alpha\beta}t^{\alpha}t^{\beta} -\dfrac{1}{2}T ^{\mbox{\ae}} t^{\alpha}t_{\alpha}\geq 0.
\end{equation}
Again, in the current case, the above condition is violated, as now we have,  
\begin{equation}
    \lb{4.41}
     T^{\mbox{\ae}}_{\alpha\beta}t^{\alpha}t^{\beta} -\dfrac{1}{2}T^{\mbox{\ae}} t^{\alpha}t_{\alpha} =- \frac{18B^2 \beta r_0^2}{\alpha(r^2-r_0^2)^2} < 0.
\end{equation}
 
In addition, the above   spacetime actually belongs to the Friedmann universe. To show this, we first introduce two new variables $\eta$ and $\bar r$ via the relations, 
\bqn
    \lb{4.45}
    t= e^{\gamma r_0\left(\eta-\eta_0\right)}, \;\;\; \bar{r} ^2=\frac{r^2}{(r^2-r_0^2)^2},
 \eqn
where $\eta_0$ is a constant, and $\gamma=\sqrt{{12}/{\alpha}}$.  Then, in terms of $\eta$ and $\bar r$, the above metric takes the form, 
\begin{equation}
    \lb{4.51}
    ds^2=a(\eta)^2\left(-d\eta^2+\frac{d\bar{r}^2}{1+4r_0^2R^2}+\bar{r}^2d\Omega^2 \right),
\end{equation}
where $R \equiv a \bar r$, and 
\begin{equation}
    \lb{4.52}
    a(\eta)=\gamma r_0 \exp{(\gamma r_0(\eta-\eta_0))}.
\end{equation}

Remember that $r_0$ was an integration constant, and from Eqs.(\ref{4.27})-(\ref{4.28}) we see that we cannot set $r_0$ to zero.  If we set $r^2_0=1/4$ then the metric of Eq.(\ref{4.51}) would be the traditional form for an FLRW metric of constant negative curvature ($k=-1$).  A solution equivalent to this was also found in \cite{Chan2020}.

\section{Spherically Symmetric Spacetimes in Painlev\`e-Gullstrand Coordinates}
\lb{SecIV}
\renewcommand{\theequation}{4.\arabic{equation}} \setcounter{equation}{0}

In this section, using the gauge freedom (\ref{1.4}), we choose the gauge 
\bqn 
    \lb{5.1a}
      g_{rr}=1, \quad R(t,r)=r, 
\eqn 
so the metric takes the   Painlev\`e-Gullstrand (PG) form, 
 \begin{equation}
    \lb{5.1}
    ds^2=-e^{2\mu(t, r)}dt^2+2e^{\nu(t, r)}drdt+  dr^2 + r^2d\Omega^2. 
\end{equation} 
For this metric we only consider time-independent solutions, and assume that  the aether is comoving,
 {$u^a =e^{-\mu}\delta^a_t$}. So, the aether is aligned with the timelike Killing vector of the metric, which is itself hypersurface-orthorgonal.
To simplify the field equations, we first define the quantity $ \lambda $, 
\begin{equation}
    \lb{5.2a}
    \lambda  \equiv e^{2\mu}+e^{2\nu}.
\end{equation}
Then,  for the static spacetimes,
$\mu = \mu(r),\quad \nu = \nu(r)$, 
   the non-vanishing components of $G_{\mu\nu}$ and $ T^{\mbox{\ae}}_{\mu\nu}$ are given by Eqs.(\ref{B.26}) - (\ref{B.34}).  The aether dynamical equations are identically zero for any $\mu$ and $\nu$, and the remaining field equations are the ones given by the ($tt$, $rr$, $\theta\theta$) components,  
\bqn 
    \lb{5.3}
    &&0=e^{4\mu-2\nu}\left[c_{14}(2r^2\mu''+4r\mu'+r^2\mu'^2)\right]\nb\\
    &&~~~~+e^{2\mu}\left[c_{14}(2r^2\mu''+4r\mu'+r^2\mu'^2+2r^2\mu'(\mu'-\nu'))\right.\nb\\
    &&~~~~\left. +4r(\mu'-\nu')\right]-2 \lambda ,\\
    \lb{5.4}
    &&0=e^{2\mu}\left[c_{14}(4r\mu'+2r^2\mu''-r^2\mu'^2)-8r\mu'\right]\nb\\
    &&~~~~~~e^{2\nu}\left[c_{14}(4r\mu'+2r^2\mu''+2r^2\mu'(\mu'-\nu'))-4r\nu'\right]\nb\\
    &&~~~~~-e^{4\mu-2\nu}\left[4r\mu'+c_{14}r^2\mu'^2\right]+2 \lambda ,\\
    \lb{5.5}
    &&0= \lambda \left[(c_{14}-2)r^2\mu'^2-2r^2\mu''-2r\mu'\right]\nb\\
    &&~~~~~+e^{2\nu}\left[2r\nu'(1+r\mu')-2r\mu'-2r^2\mu'^2\right].
\eqn 
As can be shown from  Eqs.(\ref{B.26})-(\ref{B.34}), the Einstein-aether equations  require that
\begin{equation}
    \lb{5.6}
    \lambda   \neq 0,
\end{equation}
although this is not evident from the field equations.  So,  as we proceed we must reject outright any solution that violates Eq.(\ref{5.6}).

Our strategy is to first solve the $tt$ equation for $\nu'$.  The result is
\bqn
    \lb{5.7}
    &&\nu'=\frac{-2e^{4\nu}+c_{14}(4r\mu'+r^2\mu'^2+2r^2\mu'')e^{4\mu}}{2re^{2\mu+2\nu}(2+c_{14}r\mu')}\nb\\
    &&~~~~+\frac{e^{2\mu+2\nu}(c_{14}(r\mu'(4+3r\mu')+2r^2\mu''))}{2re^{2\mu+2\nu}(2+c_{14}r\mu')}\nb\\
    &&~~~~+\frac{e^{2\mu+2\nu}(4r\mu'-2)}{2re^{2\mu+2\nu}(2+c_{14}r\mu')}.
\eqn
Note that in deriving the above expression, we assume that  \bqn 
    \lb{5.8}
    r\mu'\neq -\frac{2}{c_{14}}.
\eqn
 When $2+c_{14}r\mu'=0$, the solutions are different. So, let us pause here for a while, and first consider the case $2+c_{14}r\mu'=0$.
 
\subsubsection{$2+c_{14}r\mu'=0$}

In this case from the $tt$ equation we  find, 
\begin{equation}
    \lb{5.11}
    \mu=\frac{2}{c_{14}}\ln{\left(\frac{r_0}{r}\right)},
\end{equation}
where $r_0$ is the integration constant. By substituting this into the $rr$ equation we find, 
\begin{equation}
    \lb{5.12}
    (4 - c_{14})e^{4\nu}+2\left(\frac{r_0}{r}\right)^{\frac{8}{c_{14}}} = (c_{14}-6)e^{2\nu}\left(\frac{r_0}{r} \right)^{\frac{4}{c_{14}}},
\end{equation}
which has two  solutions, but one does not satisfy Eq.(\ref{5.6}), so we must reject it. Then, we have,  
\bqn 
     \lb{5.14}
    &&e^{2\nu}=\frac{2}{c_{14}-4}\left(\frac{r_0}{r} \right)^{\frac{4}{c_{14}}}.
\eqn  
Upon substituting Eqs.(\ref{5.11}) and (\ref{5.14}) into the  $tt$ and $\theta\theta$ field equations, we find that  
\begin{equation}
    \lb{5.15}
    c_{14}=2. 
\end{equation}
  Unfortunately this leads the solution of Eq.(\ref{5.14}) to  violate Eq.(\ref{5.6}).  It is also unphysical, as it strongly violates the constraints (\ref{2.16}),  so in the case
   $2+c_{14}r\mu'=0$ no physically acceptable solutions exist.
  
\subsubsection{$2+c_{14}r\mu'\neq 0$}

This is the case in which Eq.(\ref{5.7}) holds.  We substitute the value for $\nu'$ from this equation into the $rr$ equation and solve for $e^{2\nu}$.  The result is, 
\begin{equation}
    \lb{5.16}
    e^{2\nu}=e^{2\mu}\left(2r\mu'+\frac{c_{14}}{2}r^2\mu'^2 \right).
\end{equation}
We can substitute this value into Eq.(\ref{5.7}), and then obtain the expressions for both $\nu'$ and $e^{2\nu}$ in terms of $\mu$ and its derivatives.  In particular, 
we find,  
\bqn 
    \lb{5.16a}
    \nu'&=& \frac{c_{14}\mu''(2+r\mu'(4+c_{14}r\mu'))}{\mu'(2+c_{14}r\mu')(4+c_{14}r\mu')}\nb\\
    && +\frac{4(c_{14}-1)+c_{14}r\mu'(2+r\mu')(4+c_{14}r\mu')}{r(2+c_{14}r\mu')(4+c_{14}r\mu')}. ~~~~ 
\eqn 
Substituting it   into either the $tt$ or $rr$ equation, we find, 
\bqn
    \lb{5.17}
    &&0=(c_{14}-2)\left(4\mu'+4r\mu'^2+c_{14}r^2\mu'^3+2r\mu''\right)\nb\\
    &&\times \left[8+2(8+c_{14})r\mu'+8c_{14}r^2\mu'^2+c_{14}^2r^3\mu'^3\right].
\eqn 
So,  there exist three possibilities, 
\bqn 
    \lb{5.18}
    &&0=c_{14}-2,\\
    \lb{5.19}
    &&0=8+2(8+c_{14})r\mu'+8c_{14}r^2\mu'^2+c_{14}^2r^3\mu'^3,\\
    \lb{5.20}
    &&0=4\mu'+4r\mu'^2+c_{14}r^2\mu'^3+2r\mu''.
\eqn 
The case of  Eq.(\ref{5.18}) is not only   unphysical but also violates the constraint  (\ref{5.7}).  Therefore, 
in the following we only need to consider the last two cases. 
 
Let us first consider Eq.(\ref{5.19}),  which can be written in the form,  
\begin{equation}
    \lb{5.22}
    0=(f-\beta_0)(f-\beta_1)(f-\beta_2),
\end{equation}
Where now $ f \equiv r\mu'$, and  
\bqn 
    \lb{5.23}
    &&\beta_0=-\frac{4}{c_{14}},\nb\\
    \lb{5.24}
    &&\beta_1=-\frac{2+\sqrt{4-2c_{14}}}{c_{14}},\nb\\
    \lb{5.25}
    &&\beta_2=\frac{-2+\sqrt{4-2c_{14}}}{c_{14}}.
\eqn 
Generically, the solution to each   case is of the form, 
\bqn
    \lb{5.26}
    \mu=\ln{\left(\frac{r}{r_0} \right)^{\beta_i}},\quad
    e^{2\mu}=\left(\frac{r}{r_0} \right)^{2\beta_i},
\eqn
where $\beta_i$ is any of the ones given in  Eq.(\ref{5.23}).  When we insert Eq.(\ref{5.26}) into Eq.(\ref{5.16}) we find that, 
\begin{equation}
    \lb{5.28}
    e^{2\nu}=\frac{1}{2}\beta_i(4+c_{14}\beta_i)\left(\frac{r}{r_0} \right)^{2\beta_i}.
\end{equation}
Since any solution in which $e^{2\nu}=0$ is equivalent to the Minkowski metric, we ignore the case of $\beta_0$, as this would make $e^{2\nu}=0$, as can be seen from Eq.(\ref{5.28}).  If we insert the others $\beta_i$    
 into Eq.(\ref{5.28}), then we have, 
\begin{equation}
    \lb{5.29}
    e^{2\nu}=-\left(\frac{r}{r_0} \right)^{2\beta_i}.
\end{equation}
Unfortunately this violates the constraint of Eq.(\ref{5.7}), so we must reject it, and assume that Eq.(\ref{5.19}) does not hold.

This brings us to Eq.(\ref{5.20}), which we rewrite it as, 
\begin{equation}
    \lb{5.30}
    0=4r\mu'4r^2\mu'^2+c_{14}r^3\mu'^3+2r^2\mu''.
\end{equation}
From $f \equiv r\mu'$  we find, 
\begin{equation}
    \lb{5.31}
    r^2\mu''=rf'-f,
\end{equation}
and thus we can rewrite Eq.(\ref{5.30}) as
\begin{equation}
    \lb{5.32}
    f'=-\frac{f}{r}\left(1+2f+\frac{c_{14}}{2}f^2 \right).
\end{equation}
But this is precisely the same equation as for the static case in the Schwarzschild coordinates, given by   Eq.(26) of \cite{Eling2006-1}.  Then, we can  find the corresponding solutions  by proceeding exactly in the same way as done in
\cite{Eling2006-1}.  In particular, the solution for $\mu$ is given by,   
\begin{equation}
\lb{5.33}
    \mu(f)=\ln\left[\left( f_0\frac{1-f/f_-}{1-f/f_+} \right)^{\frac{f_+f_-}{f_+-f_-}}\right],
\end{equation}
where 
\begin{equation}
\lb{5.34}
    f_\pm=\frac{-1\pm\sqrt{1-\alpha}}{\alpha},
\end{equation}
and $f= f(r)$ is given implicitly via the relation, 
\begin{equation}
\lb{5.35}
    \frac{r_0}{r}=\left(\frac{f}{f-f_-} \right) \left(\frac{f-f_-}{f-f_+} \right)^{\frac{1}{2(1+f_+)}}.
\end{equation}

Considering the fact that the coordinate transformations,
\bq
\lb{5.35a}
\bar t= t + g(r),
\eq
can bring the PG metric to the Schwarzschild one by properly choosing the function $g(r)$, we find that in such coordinate systems  we do have $g_{tt} = g_{\bar t \bar t}$. So, the above solutions should be the ones
 found in \cite{Eling2006-1}, but written in the 
PG coordinates.

\section{Spacetimes in the Schwarzschild Coordinates}
\lb{SecV}
\renewcommand{\theequation}{5.\arabic{equation}} \setcounter{equation}{0}

The Schwarzschild  coordinates correspond to the choice, 
\bqn 
    \lb{3.0a}
    g_{tr}=0,\quad R(t,r) = r,
\eqn
for which the metric takes the form, 
\begin{equation}
    \lb{3.1}
    ds^2=-e^{2\mu(t,r)}dt^2+e^{2\nu(t,r)}dr^2+r^2 d^2\Omega,
\end{equation}
and  the comoving aether vector field takes the form, 
 {$u^a=e^{-\mu}\delta^a_t$}.
 For the sake of the simplicity, we first define the quantities, 
\begin{align}
\lb{3.3}
    Q&\equiv \frac{\mu'^2}{2}-\mu'\nu'+\mu'',\\
    \lb{A.2}
    H&\equiv \frac{\dot{\nu}^2}{2} -\dot{\mu}\dot{\nu}+\ddot{\nu}.
\end{align}
Then, the non-vanishing equation for the aether dynamics is, 
\begin{equation}
\lb{3.4}
    0=(2c_{13}-(c_2+c_{13}-c_{14})r\mu')\dot{\nu}+r(c_{123}\dot{\nu'}-c_{14}\dot{\mu'}).
\end{equation}
The non-vanishing Einstein-aether vacuum equations $G_{ab}=T^{\mbox{\ae}}_{ab}$ are the ($tt$, $tr$, $rr$, $\theta\theta$) components, given,  respectively by, 
\bqn
\lb{3.5}
    &&0=e^{2\mu}\left[c_{14}Q+2c_{14}\frac{\mu'}{r}-\frac{2\nu'}{r}+\frac{1}{r^2} \right]\nb\\ 
    &&~~~~~~~-e^{2\nu}\frac{c_{123}}{2}\dot{\nu}^2 -\frac{e^{2(\mu+\nu)}}{r^2},\\
    \lb{3.6}
    &&0=c_{14}(\dot{\mu'}-\mu'\dot{\nu})-\frac{2\dot{\nu}}{r}, \\
    \lb{3.7}
    &&0=e^{2\nu}\left[c_{123} H\right]+\frac{e^{2\nu+2\mu}}{r^2}\nb\\ 
    &&~~~~~~~-e^{2\mu}\left[\frac{2\mu'}{r}+\frac{1}{r^2}+\frac{c_{14}}{2}\mu'^2 \right], \\
    \lb{3.8}
    &&0=e^{2\mu}\left[\frac{\mu'^2}{2}\left(c_{14}-1 \right)+\frac{\nu'-\mu'}{r}-Q \right]\nb\\ 
    &&~~~~~~~+e^{2\nu}\left[\frac{\dot{\nu}^2}{2}\left(1-c_{13} \right) +\left(c_2+1 \right)H \right].
\eqn

\subsection{Time-independent Solutions}

The static solution was already found in \cite{Eling2006-1} but with  different (though equivalent) parameterizations of the metric.
  In the static case, all time-derivatives go to zero and the ($tt$, $rr$, $\theta\theta$) equations  become, 
\begin{align}
\lb{3.9}
    \frac{e^{2\nu}}{r^2}&=c_{14} Q+2c_{14}\frac{\mu'}{r}-2\frac{\nu'}{r}+\frac{1}{r^2}, \\
    \lb{3.10}
    \frac{e^{2\nu}}{r^2}&=c_{14}\frac{\mu'^2}{2}+2\frac{\mu'}{r}+\frac{1}{r^2}, \\
    \lb{3.11}
    0&=\frac{\mu'^2}{2}(c_{14}-1)+c_{14}\mu''-c_{14}\mu'\nu'.
\end{align}

Subtracting the $rr$ equation from the $tt$ one, we   find, 
\begin{equation}
\lb{3.12}
    2\frac{\nu'}{r}=c_{14}\mu''-c_{14}\mu'\nu'+2\frac{\mu'}{r}(c_{14}-1).
\end{equation}
On the other hand, from the $\theta\theta$ equation we obtain,  
\begin{equation}
\lb{3.13}
    2\frac{\nu'}{r}=2\mu''+\mu'^2(2-c_{14})+2\frac{\mu'}{r}-2\mu'\nu',
\end{equation}
which, together with eq.(\ref{3.12}) yields, 
\begin{equation}
\lb{3.15}
    \nu'=\frac{\mu''}{\mu'}+\mu'+\frac{2}{r}.
\end{equation}
 We can rewrite the $\theta\theta$ equation as, 
\begin{equation}
\lb{3.16}
    \nu'\left( \frac{1}{r}+\mu' \right)=\frac{\beta}{2}\mu'^2+\frac{\mu'}{r}+\mu''.
\end{equation}
Inserting our expression (\ref{3.15}) into this equation and after simplification,  we find
\begin{equation}
\lb{3.17}
    r^2\mu''+2r\mu'+2r^2\mu'^2+\frac{c_{14}}{2}r^3\mu'^3=0,
\end{equation}
which is equivalent to Eq.(26) of \cite{Eling2006-1}, provided that we make the following substitutions, 
\bq
\lb{3.18}
    c_{14}\rightarrow c_1, \;\;\;
    \mu\rightarrow \frac{A}{2}.
\eq
We can solve this using an equivalent process as that given in \cite{Eling2006-1}.  In particular, setting 
$f=r\mu'$, 
we find that  Eq.(\ref{3.17}) becomes
\begin{equation}
\lb{3.20}
    \frac{df}{dr}=-\frac{f}{r}\left(1+2f+\alpha f^2 \right),
\end{equation}
but now with $\alpha\equiv c_{14}/2$.  
From the chain role, $ \frac{d\mu}{dr}=\frac{d\mu}{df}\frac{df}{dr}$,
 and the definition of $f$ we find
\begin{equation}
\lb{3.22}
    \frac{d\mu}{df}=-\dfrac{1}{1+2f+\alpha f^2}.
\end{equation}
Using the partial fraction decomposition we can solve the above equation, and find
\begin{equation}
\lb{3.23}
    \mu(f)=\ln\left[\left( f_0\frac{1-f/f_-}{1-f/f_+} \right)^{\frac{f_+f_-}{f_+-f_-}}\right],
\end{equation}
where $f_0$ is an integration constant whose square is unity.
The equivalent equation in \cite{Eling2006-1} is Eq.(34), and this solution matches it exactly, bearing in mind that, 
\begin{equation}
\lb{3.24}
    f_\pm=\frac{-1\pm\sqrt{1-\alpha}}{\alpha}.
\end{equation}
Then we can solve Eq.(\ref{3.20}) and   find, 
\begin{equation}
\lb{3.25}
    \frac{r_0}{r}=\left(\frac{f}{f-f_-} \right) \left(\frac{f-f_-}{f-f_+} \right)^{\frac{1}{2(1+f_+)}},
\end{equation}
which is equivalent to Eq.(35) of \cite{Eling2006-1}.

Note that, when   $c_{14}=2$,   instead of Eq.(\ref{3.23}) now we have
\begin{equation}
    \lb{3.25a}
    \mu(f)=\ln\left[\left( f_0\frac{f+f_+}{f+f_-} \right)^{\frac{1}{f_--f_+}}\right],
\end{equation}
where now $f_\pm$ are defined by
\begin{equation}
    \lb{3.25b}
    f_\pm=\dfrac{3}{4}\pm\dfrac{\sqrt{41}}{4},
\end{equation}
and instead of Eq.(\ref{3.25}) we   have
\begin{equation}
    \lb{3.25c}
    \frac{r_0}{r}=f^{\frac{2}{f_+f_-}}\left(\frac{(f-f_+)^{1/f_+}}{(f-f_-)^{1/f_-}} \right)^{\frac{2}{f_+-f_-}}.
\end{equation}
But, as shown above, this solution is physically not acceptable.

\subsection{Time-dependent Solutions}

If we consider solutions such that $e^\mu$ and $e^\nu$ are separable in $t$ and $r$, then we seek solutions of the forms of
Eqs.(\ref{4.11}) and  (\ref{4.12}).  
In this case,   the $tr$ and aether equations reduce to, 
\bqn
    \lb{3.30a}
    &&\frac{2}{r}=-c_{14}\mu',\\
    \lb{3.31}
    &&\frac{2c_{13}}{r}=\mu'(c_{14}-c_{123}).
\eqn
We now consider separately the cases $c_{13}=0$ and $c_{13} \not=0$.

\subsubsection{$c_{13}=0$}

By Eq.(\ref{3.30a}) we must have
\bqn
c_{14}\neq 0, \quad
\mu'\neq 0.
\eqn
Then, from Eq.(\ref{3.31}) we have, 
\bq 
    \lb{3.32}
    c_2=c_{14},
\eq 
for which   Eq.(\ref{3.30a}) yields, 
\bq
    \lb{3.33}
    \mu=\ln{\left(\frac{U_0}{r^\alpha}\right)},\;\;\;  \alpha\equiv\frac{2}{c_2},
\eq
where $U_0$ is an arbitrary constant. Then, the $tt$, $rr$, $\theta\theta$  equations (\ref{3.5}), (\ref{3.7}) and (\ref{3.8}) become
\bqn
    \lb{3.35}
    &&\dfrac{U_0^2}{r^{2\alpha+2}}\left(\alpha-1 \right)=e^{2\nu}\left[\dfrac{\dot{\nu}^2}{\alpha}+\dfrac{U_0^2}{r^{2\alpha+2}}\right],\\
    \lb{3.36}
    &&\dfrac{U_0^2}{r^{2\alpha+2}}\left(\alpha-1 \right)=e^{2\nu}\left[-\dfrac{2}{\alpha}\ddot{\nu}-\dfrac{\dot{\nu}^2}{\alpha}-\dfrac{U_0^2}{r^{2\alpha+2}}\right],\\
    \lb{3.37}
    &&\dfrac{U_0^2}{r^{2\alpha+2}}\left(\alpha-1 \right)\left(r\nu'+\alpha \right)=e^{2\nu}\Bigg[\dfrac{2}{\alpha}\ddot{\nu}+\dfrac{\dot{\nu}^2}{\alpha}\nb\\
    &&~~~~~~~~~~~~~~~~~~~~~~~~~~~~~~~ +\ddot{\nu}+\dot{\nu}^2 \Bigg].
\eqn
By combining the $tt$ and $rr$ equations we find
\bq
    \lb{3.38}
    \dfrac{\dot{\nu_1}(t)^2+\ddot{\nu_1}(t)}{\alpha}=-\dfrac{U_0^2}{r^{2\alpha+2}},
\eq 
where we have explicitly written the expressions for $\dot{\nu}$ in terms of $\nu_1(t)$ to emphasize the $t$-dependence.  Since the left-hand side (LHS) is purely $t$-dependent, and the right-hand side (RHS) is purely $r$-dependent, then both sides must be equal to some constant.  Since   $U_0 \not=0$,  the only way to ensure that the RHS of Eq.(\ref{3.38}) is constant is to set, 
\bqn
    \lb{3.39}
    \alpha=-1,
\eqn
for  which   Eq.(\ref{3.38}) reduces to
\bq
    \lb{3.41}
    \dot{\nu}^2+\ddot{\nu}=U_0^2.
\eq 
By using Eq.(\ref{3.41}) with either the $tt$ or $rr$ equation, we arrive at
\begin{equation}
    \lb{3.42}
    2U_0^2=e^{2\nu_1}\left(\dot{\nu}_1^2-U_0^2\right), 
\end{equation}
which yields, 
\begin{equation}
    \lb{3.45}
    \nu_1(t)=\ln{\left\{\sqrt{2}\sinh\left[U_0\left(t_0 \pm t\right)\right]\right\}},
\end{equation}
where $t_0$ is an arbitrary constant. On the other hand,  from Eq.(\ref{3.37}) we find
\begin{equation}
    \lb{3.46}
    e^{2\nu}\left(\dot{\nu}^2-U_0^2\right) =2 U_0^2(1-r\nu').
\end{equation}
Comparing this to Eq.(\ref{3.42}),  we find $   \nu_0(r)=$ const., so  that  
\begin{equation}
    \lb{3.48}
    \nu(t,r)=\ln{\left\{\sqrt{2}\sinh\left[U_0\left(t_0 \pm  t\right)\right]+V_0\right\}},
\end{equation}
where $V_0$ is a constant. Eqs.(\ref{3.33}) and (\ref{3.48}) satisfy all of the field equations, provided that 
$V_0=0$,  with no other constraints on the remaining  arbitrary constants.
Thus for the case $c_{13} = 0$,  the solution   is
\bqn
    \lb{3.49}
    &&\mu(r)=\ln{\left(U_0 r\right)},\nb\\
    &&\nu(t,r)=\ln{\left[\sqrt{2}\sinh\left(U_0(t_0 \pm t)\right)\right]}.
\eqn
However, using the gauge freedom for the choice of $t$, we can always set $U_0 = 1$ and $t_0 = 0$, so the metric finally takes the form,
\begin{equation}
    \lb{3.54}
    ds^2=-r^2 dt^2+ 2 \sinh^2\left( t\right)dr^2 +r^2 d^2\Omega.
\end{equation}

Unfortunately, this solution is also excluded by the current observations, as Eqs.(\ref{3.32}), (\ref{3.33}) and (\ref{3.39}) imply that 
\bq
\lb{3.54aa}
c_{14} = c_2 = -2. 
\eq

\subsubsection{$c_{13}\neq 0$}

In this case we combine the $tt$ and $rr$ equations, and find
\begin{equation}
    \lb{3.54a}
    c_{123}(\dot{\nu_1}^2+\ddot{\nu_1})=-\frac{2 U_0^2}{r^{2\alpha+2}}.
\end{equation}
 Substituting it into the $tt$ equation, and then subtracting it from the $\theta \theta$ equation, we find obtain, 
\begin{equation}
    \lb{3.54b}
    2 U_0^2 r \nu_0'(r)e^{-2\nu_0(r)}=e^{2\nu_1(t)}\left[\frac{2 c_{13}}{1+c_{13}} U_0^2 \right].
\end{equation}
The right-hand is always different from zero, so the above equation holds only when $\nu_1 = $ const. Then, 
 {from Eq.(\ref{3.54a}) we find that the integration constant $U_0$ must vanish. As a result,
Eq.(\ref{3.54b}) becomes an identity ($0=0$). It can be shown that the rest of the Einstein-aether field equations will
give the static solutions presented in the last subsection.}

\section{Conclusions}
\lb{SecVI}
\renewcommand{\theequation}{6.\arabic{equation}} \setcounter{equation}{0}
 
With the increasing interest of Einstein-aether theory in the recent years, in this paper we have studied spherically symmetric both static and time-dependent spacetimes in this theory, and
found several exact solutions in closed forms.  
Such studies were carried out in three different coordinate systems:  {\em the isotropic,  
Painlev\`e-Gullstrand, and Schwarzschild coordinates}, and in each of them exact solutions are found. 

In particular, in the isotropic coordinates we have found a class of exact static solutions  in closed forms, given by Eq.(\ref{4.static24}), i.e., 
\bqn
    \lb{6.1}
    ds^2&=&-\left(\frac{1-\frac{m}{2r}}{1+\frac{m}{2r}} \right)^qdt^2 + \frac{\left(1+\frac{m}{2r}\right)^{q +2}}{\left(1+\frac{m}{2r}\right)^{q -2}}\nb\\
     &&  
     \times \Big[dr^2 + r^2 \left(d^2\theta + \sin^2\theta d^2\phi\right)\Big], ~~~~
\eqn
where 
\bq
\lb{6.2}
q \equiv 2 \left(\frac{2}{2-c_{14}}\right)^{1/2}. 
\eq
Clearly, when $c_{14} = 0$, the above solution reduces to the Schwarzschild vacuum black 
hole solution but written in the isotropic coordinates, and the spacetime is free of spacetime curvature singularities at  $r = m/2$ \cite{D'Inv}. 

However,  as long as $c_{14}\not= 0$ but satisfies  the observational constraint (\ref{2.16}), i.e., 
\bq
\lb{6.3}
0 < c_{14} \leq 2.5\times 10^{-5},
\eq
 the corresponding spacetime has several remarkable features:
 \begin{itemize}
 
 \item A minimal surface with non-zero area always exists at $\bar{r}=\bar{r}_{\text{min}}$, given explicitly by  Eq.(\ref{Static_4.static36}), the so-called throat of the spacetime. 
 It smoothly connects two regions, $\bar{r} \in (2m, \bar{r}_{\text{min}}]$ and $\bar{r} \in [\bar{r}_{\text{min}}, \infty)$, as schematically shown by Fig. \ref{fig1}.
 
 \item The Kretschmann scalar always diverges  at $\bar{r} = 2m$ as long as $c_{14} \not= 0$, so a spacetime curvature singularity always appears.
 Despite the fact $A(\bar{r} = 2m) = \infty$, the proper radial distance between the throat and the singularity is always finite and non-zero [cf.  Eq.(\ref{Static_4.static36b})].

 \item In the region $\bar{r} \in [\bar{r}_{\text{min}}, \infty)$, the spacetime is asymptotically flat as $\bar{r} \rightarrow \infty$, and the proper radial distance 
 between the throat and the spatial infinity $\bar{r} = \infty$ is always infinitely large, so is the geometric area, $A(\bar{r} = \infty) = \infty$.

 \item The throat is only marginally trapped, as now  $\Theta_{+}\Theta_{-}$ vanishes precisely only at the throat, $ \bar{r}  = \bar{r}_{\text{min}}$, 
 while away from it, we always have $\Theta_{+}\Theta_{-} < 0$,
 as shown explicitly by Eq.(\ref{Static_4.static38}), where $\Theta_{\pm}$ denote the expansions of the outgoing/ingoing null geodesic congruences.
    
  \end{itemize}
 
 With these remarkable features, it would be very interesting to consider other properties of the solution, including its  stability  against non-spherical perturbations
  and consistency with Solar System tests \cite{ZW20} and  the observations of the shadows of black holes \cite{AA20}. 
   {In particular, in general relativity, in order to have
  a throat that connects two un-trapped regions, exotic matter (for example, the one that violates  energy conditions)  is often requested, which indicates some kind of  instabilities \cite{MV96}.
  In the present case, although such a kind of matter is provided by the aether field, it would be very interesting to show the stability of the throat against perturbations, 
  especially against the non-spherical ones.}
  
  { It would also be important to study the corresponding solutions in
the context of the UV completion of the Einstein-aether theory, i.e.
the non-projectable Ho\v{r}ava gravity. In particular it is intriguing
to see whether and how the naked singularity present in the case of
$q\ne 2$ could be resolved or hidden behind a universal horizon.}
 
 In addition,  to simplify mathematically the problems involved, in this paper we have considered only the cases in which the aether field is always comoving with the chosen coordinate systems. In general,
 the aether field can have radial motions, as long as it is timelike. It would be very  interesting, if exact solutions with closed forms can be found in this case, too.

\section*{Acknowledgements}

 {We would like to express our gratitude to Ted Jacobson for carefully reading our manuscript and valuable suggestions and comments.}
This work was partially supported by the National Key Research and Development Program of China under  the Grant No. 2020YFC2201503, 
the National Natural Science Foundation of China under the grant Nos.   11975203,
the Japan Society for the Promotion of Science Grants-in-Aid for Scientific Research No. 17H02890, 
 No. 17H06359, and the World Premier International Research Center Initiative, MEXT, Japan.

\section*{Appendix A: $G_{\mu\nu}$ and $T^{\text{\ae}}_{\mu\nu}$ in different Coordinate Systems}
\renewcommand{\theequation}{A.\arabic{equation}} \setcounter{equation}{0}

In this appendix, we shall present the non-vanishing components of the Einstein tensor $G_{\mu\nu}$ and the effective energy-momentum tensor $T^{\text{\ae}}_{\mu\nu}$
in three different coordinate systems. In all of them, the aether is assumed to be comoving with the coordinate systems, so we always have
\begin{equation}
    \lb{A.0a}
    u^{\mu}= \pm \left(-g_{tt}\right)^{-1/2} \lambda ^{\mu}_t,
\end{equation}
where the ``+'' sign means the aether field is moving along $dt$ increasing direction, while  the ``-'' sign corresponds to the case in which 
 the aether field is moving along $dt$ decreasing direction. From Eqs.(\ref{2.6}) and (\ref{2.7}), we can see that  $T^{\text{\ae}}_{\mu\nu}$ is independent of these chocies,
 while and $\text{\AE}$   switches its sign. Clearly, these choices do not affect the field equations (\ref{2.1}) and (\ref{2.7}). With these in mind, let us consider the three different 
 coordinate systems.

\subsection{Isotropic  Coordinates}

Choosing the gauge (\ref{4.0a}), the metric takes the form, 
\begin{equation}
    \lb{A.0a2}
    ds^2=-e^{2\mu(r,t)}dt^2+e^{2\nu(t,r)}\left(dr^2+r^2 d^2\Omega\right),
\end{equation}
Where $d^2\Omega \equiv d\theta^2+\sin{\theta}^2d\phi^2$. 
Introducing the quantity, 
\begin{equation}
    \lb{B.1}
    \Sigma=3\dot{\nu}^2+2\ddot{\nu}-2\dot{\mu}\dot{\nu}, 
\end{equation}
we find that  the non-zero and independent components of the Einstein tensor are, 
\bqn
\lb{B.2}
&&G_{00}=3\dot{\nu}^2-e^{2\mu-2\nu}\left(\nu'^2+2\nu''+4\frac{\nu'}{r} \right),\\
\lb{B.3}
&&G_{01}=2\mu'\dot{\nu}-2\dot{\nu}',\\
\lb{B.4}
&&G_{11}=\nu'^2+2\mu'\nu'+\frac{2}{r}(\mu'+\nu')-\Sigma e^{-2\mu+2\nu},  \\
\lb{B.5}
&&G_{22}=r^2\Big(\mu'^2+\mu''+\nu''+\frac{\mu'+\nu'}{r}-\Sigma e^{-2\mu+2\nu}\Big),~~~~~
\eqn
and the non-zero and independent components of the aether stress-energy tensor are, 
\bqn
\lb{B.8}
    &&T^{\mbox{\ae}}_{00}= e^{2\mu-2\nu}\Bigg[c_{14}\Big(\frac{\mu'^2}{2}+\mu'\nu'+\mu''+2\frac{\mu'}{r}\Big)\Bigg]\nb\\
    &&~~~~~~~~~~~~~~~~~~~~ -\frac{3}{2}\beta\dot{\nu}^2,    \\
    \lb{B.9}
    &&T^{\mbox{\ae}}_{01}=c_{14}\left(\dot{\mu}'+\mu'\dot{\nu} \right),  \\
    \lb{B.10}
    &&T^{\mbox{\ae}}_{11}=\dfrac{\beta}{2}\Sigma e^{-2\mu+2\nu} -\frac{c_{14}}{2}\mu'^2 , \\
    \lb{B.11}
    &&T^{\mbox{\ae}}_{22}=r^2\left(T_{11}^{\mbox{\ae}}+c_{14}\mu'^2\right),    
\eqn
where $ \beta\equiv 3c_2+c_{13}$.

\subsection{Painlev\`e-Gullstrand Coordinates}

Choosing the gauge (\ref{5.1a}),   and considering only the static spacetimes, we find that 
the metric takes the   Painlev\`e-Gullstrand (PG) form, 
 \begin{equation}
    \lb{B.24a}
    ds^2=-e^{2\mu(r)}dt^2+2e^{\nu(r)}drdt+  dr^2 + r^2d\Omega^2.  
\end{equation} 
Setting
\begin{equation}
    \lb{B.25}
    \lambda=e^{2\mu}+e^{2\nu},
\end{equation}
we find that the  non-zero components of the Einstein tensor are, 
\bqn 
    \lb{B.26}
    &&G_{00}=\frac{1}{\lambda^2r^2}\Big[e^{4\mu+2\nu}\left(1-2r\left(\mu'+\nu'\right)\right)\nb\\
    &&~~~~~~~~~~~+e^{2\mu+4\nu}\Big],\\
    \lb{B.27}
    &&G_{01}=\frac{1}{\lambda^2r^2}\Big[e^{2\mu+3\nu}\left(-1+2r\left(\mu'-\nu'\right)\right)\nb\\
    &&~~~~~~~~~~~-e^{5\nu}\Big],\\
    \lb{B.28}
    &&G_{11}=\frac{1}{\lambda^2r^2}\Big[e^{2\mu+2\nu}\left(4r\mu'-1\right)+2e^{4\mu}r\mu'\nb\\
    &&~~~~~~~~~~~+e^{4\nu}\left(2r\nu'-1\right)\Big],\\
    \lb{B.29}
    &&G_{22}=\frac{1}{\lambda^2r^2}\Bigg[e^{4\mu}\left(r\mu'+r^2\mu'^2+r^2\mu''\right)\nb\\
    &&~~~~~~~~~~+e^{2\mu+2\nu}\Big(r\left(1+r\mu'\right)\left(2\mu'-\nu'\right)\nb\\
    &&~~~~~~~~~~+r^2\mu''\Big)\Bigg],
\eqn 
while the non-zero components of the aether stress-energy tensor are, 
\bqn
\lb{B.31}
    &&T^{\mbox{\ae}}_{00}=\frac{c_{14}}{2\lambda^2r^2}\Bigg[e^{6\mu}\left(4\mu'+r\mu'^2+2r\mu''\right)\nb\\
    &&~~~~~~~~+e^{4\mu+2\nu}\Big(4\mu'+3r\mu'^2-2r\mu'\nu'\nb\\
    &&~~~~~~~~+2r\mu''\Big)\Bigg],    \\
    \lb{B.32}
    &&T^{\mbox{\ae}}_{01}=\frac{c_{14}}{2\lambda^2r^2}\Bigg[e^{2\mu+3\nu}\left(4\mu'+3r\mu'^2+2r\mu''-2r\mu'\nu'\right)\nb\\
    &&~~~~~~~~-e^{4\mu+\nu}\left(4\mu'+r\mu'^2+2r\mu'\nu'\right)\Bigg],    \\
    \lb{B.33}
    &&T^{\mbox{\ae}}_{11}=\frac{c_{14}}{2\lambda^2r^2}\Bigg[e^{4\nu}\left(2\mu'+r\mu'^2+r\mu''\right)\nb\\
    &&~~~~~~~~+e^{2\mu+2\nu}\left(4\mu'-r\mu'^2+2r\mu'\nu'\right)\nb\\
    &&~~~~~~~~-4e^{4\mu}r\mu'^2\Bigg],    \\
    \lb{B.34}
    &&T^{\mbox{\ae}}_{22}=\dfrac{c_{14}e^{2\mu}r^2\mu'^2}{2\lambda}.
    \eqn

    \subsection{Schwarzschild Coordinates}
    
The Schwarzschild  coordinates correspond to the choice (\ref{3.0a}), 
for which the metric takes the form, 
\begin{equation}
    \lb{B.34a}
    ds^2=-e^{2\mu(t,r)}dt^2+e^{2\nu(t,r)}dr^2+r^2 d^2\Omega.
\end{equation}
We also define the quantities, 
\begin{align}
\lb{B.13}
    Q&=\frac{\mu'^2}{2}-\mu'\nu'+\mu'', \\
    \lb{B.14}
    H&=\frac{\dot{\nu}^2}{2} -\dot{\mu}\dot{\nu}+\ddot{\nu}. 
\end{align}
Then, the non-zero components of the Einstein tensor are, 
\bqn
\lb{B.15}
&&G_{00}=\frac{1}{r^2}e^{2(\mu-\nu)}\left(e^{2\nu}+2r\nu' -1\right),\\
\lb{B.16}
&&G_{01}=\frac{2\dot{\nu}}{r},\\
\lb{B.17}
&&G_{11}=\frac{1}{r^2}\left(1-e^{2\nu}+2r\mu'  \right),  \\
\lb{B.18}
&&G_{22}=r^2\Bigg\{e^{-2\nu} \left[Q+\left(\frac{\mu'^2}{2}-\frac{\nu'-\mu'}{r} \right)\right]\nb\\
&&~~~~~~~~~~~~~~~~ -e^{-2\mu}\left(H+\frac{\dot{\nu}^2}{2}\right)\Bigg\}, 
\eqn
and the non-zero components of the aether stress-energy tensor are, 
\bqn
\lb{B.20}
    &&T^{\mbox{\ae}}_{00}= e^{2\mu-2\nu}c_{14}\left(Q+\frac{2\mu'}{r}\right) -\frac{c_{123}}{2}\dot{\nu}^2,    \\
    \lb{B.21}
    &&T^{\mbox{\ae}}_{01}=c_{14}\left(\dot{\mu}'-\mu'\dot{\nu}\right),  \\
    \lb{B.22}
    &&T^{\mbox{\ae}}_{11}=c_{123}H e^{-2\mu+2\nu} -\frac{c_{14}}{2}\mu'^2, \\
    \lb{B.23}
    &&T^{\mbox{\ae}}_{22}=r^2\Bigg[e^{-2\mu}\left(c_2H-\frac{c_{13}}{2}\dot{\nu}^2\right)\nb\\
    &&~~~~~~~~~~~~~~~ - \frac{c_{14}}{2} e^{-2\nu}\mu'^2\Bigg].
\eqn


\end{document}